\documentclass[pdflatex,sn-nature]{sn-jnl}

\usepackage{graphicx}%
\usepackage{multirow}%
\usepackage{amsmath,amssymb,amsfonts}%
\usepackage{amsthm}%
\usepackage{mathrsfs}%
\usepackage[title]{appendix}%
\usepackage{xcolor}%
\usepackage{textcomp}%
\usepackage{manyfoot}%
\usepackage{booktabs}%
\usepackage{algorithm}%
\usepackage{algorithmicx}%
\usepackage{algpseudocode}%
\usepackage{listings}%

\setcitestyle{super,open={},close={}}

\makeatletter
\AtBeginDocument{%
  \renewcommand{\@biblabel}[1]{#1.}%
}
\makeatother

\theoremstyle{thmstyleone}%
\theoremstyle{thmstyletwo}%

\theoremstyle{thmstylethree}%

\begin{document}

\title[Article Title]{Quadrupolar phase transition in superconducting lanthanum hydride}

\author*[1,2]{\fnm{Abhishek} \sur{Raghav}}\email{mwkabr1915@icloud.com}

\author[2,3]{\fnm{Kousuke} \sur{Nakano}}\email{nakano.kosuke@nims.go.jp
}

\author[1]{\fnm{Marco} \sur{Cherubini}}\email{marco.cherubini@sorbonne-universite.fr}

\author[2,4]{\fnm{Ryotaro} \sur{Arita}}\email{arita@riken.jp
}

\author*[1]{\fnm{Michele} \sur{Casula}}\email{michele.casula@upmc.fr}

\affil*[1]{\orgdiv{Institut de Minéralogie, de Physique des Matériaux et de Cosmochimie (IMPMC)}, \orgname{Sorbonne Université CNRS UMR 7590}, \orgaddress{\street{4 Place Jussieu}, \city{Paris}, \postcode{75252}, \country{France}}}

\affil[2]{\orgdiv{Center for Emergent Matter Science}, \orgname{RIKEN}, \orgaddress{\street{2-1 Hirosawa}, \city{Wako}, \postcode{351-0198}, \state{Saitama}, \country{Japan}}}

\affil[3]{\orgdiv{Center for Basic Research on Materials}, \orgname{National Institute for Materials Science (NIMS)}, \orgaddress{\city{Tsukuba}, \postcode{305-0047}, \state{Ibaraki}, \country{Japan}}}

\affil[4]{\orgdiv{Department of Physics, The University of Tokyo}, \orgaddress{\city{Tokyo}, \postcode{113-0033}, \state{Tokyo}, \country{Japan}}}

\abstract{Lanthanum hydride (LaH{\unboldmath$_{10}$}) has been widely studied for its high superconducting critical temperature of 250 K at about 170 GPa pressure. Although the structural {\unboldmath${R\bar{3}m}$}-to-{\unboldmath${Fm\bar{3}m}$}
transition under pressure connected to the emergence of the superconducting phase in this material is broadly understood, the detailed characterization of its nature and its order parameter are still missing. 
By applying the cluster multipole moment analysis to the hydrogen sublattice, we reveal that this transition is triggered by a quadrupolar {\unboldmath$T_{2g}$} order parameter, and we provide evidence for its weak first-order nature. By performing path integral molecular dynamics coupled to a message-passing atomic cluster expansion (MACE) neural network potential, trained on Perdew–Burke–Ernzerhof (PBE) density functional theory configurations, we show that the collapse of the order parameter at the transition is simultaneously associated with the discontinuous softening of the optical {\unboldmath$T_{2g}$} phonons. Their symmetry lets them carry a non-negligible electron-phonon coupling in LaH{\unboldmath$_{10}$}, while the weak first-order nature of the transition makes them soft. 
The presence of structural instabilities with low-frequency quadrupolar distortions can be a key ingredient to enhance superconductivity in superhydrides and provides guidance for the discovery of new high-{\unboldmath$T_c$} superconductors in hydrogen-rich compounds.
}




\maketitle

\section{Introduction}\label{sec1}

Following Ashcroft's proposal of superconductivity in metallic hydrogen under extremely high pressures in 1968~\cite{1968ASH}, hydrides have been widely studied both theoretically and experimentally for high
superconducting critical temperature ($T_c$) 
applications. It was suggested that chemical precompression with heavier atoms would reduce the extreme pressure required to metallize hydrogen~\cite{2023MON} and hence, reach superconductivity at lower pressures~\cite{1971GIL,2004ASH,2022HIL}. Following this recipe and using \textit{ab initio} crystal structure 
search
methods, several hydrogen-rich compounds have been predicted to show superconductivity at pressures much lower than those necessary to metallize hydrogen, by computing 
their $T_c$ using 
McMillan or
Migdal-Eliashberg formalisms~\cite{FLORESLIVAS2020}. Several of these hydrides and superhydrides have been successfully synthesized. The most notable ones are sulfur hydride (H$_3$S, 203 K at 155 GPa)~\cite{2014YIN, 2015DRO, 2015ERR}, lanthanum hydride (LaH$_{10}$, 250 K at 170 GPa,~\cite{2019DRO,2019SOM}), and yttrium hydride (YH$_6$ and YH$_9$, 220 K at 183 GPa and 243 K at 201 GPa respectively~\cite{2017PEN,2021KON}). Interestingly, superconductivity in these hydrides was first predicted by \textit{ab initio} calculations and later followed by experimental synthesis, hence \textit{ab initio} predictions played a key role in guiding experiments. Because these are believed to be conventional BCS~\cite{1957BAR} superconductors, unlike cuprates, the mechanism of superconductivity is generally well understood, although proximity with structural instabilities could make them peculiar~\cite{2022FRA,2026CHE}. 

High-pressure hydrides have rich phase diagrams characterized by several structural transitions. In a number of hydride families, maximum $T_c$ is found in proximity to 
symmetry-lowering transitions. This is also true in the case of LaH$_{10}$, which is made of clathrate cages of H atoms, with La atoms sitting at the centers of the cages. These cages are bridged together with cubes (or distorted cubes, depending upon the specific phase) of H atoms (see Fig.~\ref{lah10_structures}a). Experimentally, a $T_c$ of $\approx$250 K was reported in face-centered-cubic (fcc) based 
LaH$_{10}$ samples, with critical pressures ranging between 137 and 218 GPa~\cite{2019DRO, 2019SOM}. This fcc-based 
structure undergoes distortion upon decompression. The symmetry of the low-pressure distorted phase, however, is still contested, with some reports claiming it to be rhombohedral ($R\bar{3}m$)~\cite{2018GEB} and others claiming it to be $C2/m$~\cite{2021SUN}. Furthermore, it was shown that the $Fm\bar{3}m$ LaH$_{10}$ survives to pressures as low as 135 GPa, below which it starts to distort and $T_c$ begins to drop sharply~\cite{2021SUN}. Early 
density functional theory
(DFT) calculations, however, found fcc-$Fm\bar{3}m$ dynamically unstable below 210 GPa~\cite{2017LIU}; nuclear quantum effects (NQEs) were later shown to stabilize it well below this pressure~\cite{2020ERR}. Thus, the nature of this transition, - including the symmetries involved, location of the phase boundary, and the role of temperature and quantum effects -, has been constantly debated~\cite{2018GEB,2021SUN, 2017LIU, 2020ERR,2022WAT,2022KEV}. 
Here, we present a detailed picture of its nature by characterizing the symmetry-breaking mechanism taking place in the hydrogen sublattice. This reveals a quadrupolar order parameter, which triggers this transition. The order parameter also lets us locate the structural transition more precisely in the theoretical phase diagram and, more importantly, suggests a connection between the quadrupolar order and the high $T_c$ of the superconducting transition occurring nearby. 

To map the pressure-temperature ($P$-$T$) phase diagram, we perform path integral molecular dynamics (PIMD) simulations coupled to a message-passing atomic cluster expansion (MACE) machine learning interatomic potential (MLIP), trained on DFT - Perdew–Burke–Ernzerhof (PBE) configurations.
Previous PIMD investigations have predominantly employed lattice-parameter characterization and space-group symmetry analysis to locate the phase-transition boundary~\cite{2022WAT, 2022KEV}. With a sufficiently fine $P$-$T$ grid, this approach may adequately resolve the phase boundary; however, it does not, in itself, provide deeper insight into the fundamental nature or underlying mechanisms of the transition. In this study, we employ two complementary methodologies to study it. On one hand, we compute anharmonic phonons, and identify their soft modes. This not only aids in tracing the phase transition boundaries but also gives evidence for its first-order nature. On the other hand, we apply a cluster multipole (CMP) moments approach~\cite{2017SUZ} to
devise the order parameter for a more comprehensive characterization of the transition.
 The order parameter encodes the symmetry breaking into a scalar quantity and provides a robust way to map the phase diagram. We show that the distortion of the H-cubes bridging the clathrate cages can be used as signature of different phases, and hence we define the order parameter over these cubes. We find that this order parameter transforms according to the $T_{2g}$ irreducible representation and it is therefore quadrupolar in character. We finally consider the implications of such quadrupolar ordering for the superconducting state.

\section{Anharmonic phonon softening}
LaH$_{10}$ is a strongly anharmonic system, with anharmonicity arising mainly from significant NQEs at low temperatures. This is evident from the fact that harmonic phonons severely underestimate the dynamic stability region of the symmetric $Fm\bar{3}m$ phase, which is stable at much lower pressures than those predicted by the harmonic approximation (see Extended data Fig.~\ref{dfpt_vs_pimd_phonons}). Hence, we compute anharmonic phonons for LaH$_{10}$ from PIMD trajectories performed at different pressures (60 GPa -- 130 GPa) and temperatures (60 K -- 220 K). The simulations were performed for the NPT ensemble using 76 beads for the lowest temperature (60 K) case, for a length of 64 ps (see Methods for more details). PIMD gives a comprehensive amount of information about the structural phase diagram. Here we compute displacement-displacement and velocity-velocity correlation functions, and we evaluate phonons directly from the nuclear quantum distribution. This is done by solving a generalized eigenvalue equation using zero-time Kubo-transformed autocorrelation matrices built from bead-averaged (centroid) displacements and velocities~\cite{2021MOR}. For a standard pressure-driven second-order displacive transition, one would observe that the frequency of soft phonon modes approaches zero as the system undergoes the transition from the high-symmetry phase, reaching zero at the critical pressure ($P_c$) where the lattice becomes unstable 
according to the soft-mode eigenvectors. In the low-symmetry phase, below $P_c$, the phonon begins to re-harden. In a characteristic first-order transition, on the other hand, the 
phonon modes
never reach zero, but hit a finite floor, and then jump discontinuously as the system settles in a different phase. Hence, phonons serve as an effective tool to probe not only the phase transition boundary but also its character.

\begin{figure}[H]
\centering
\includegraphics[width=1.0\textwidth]{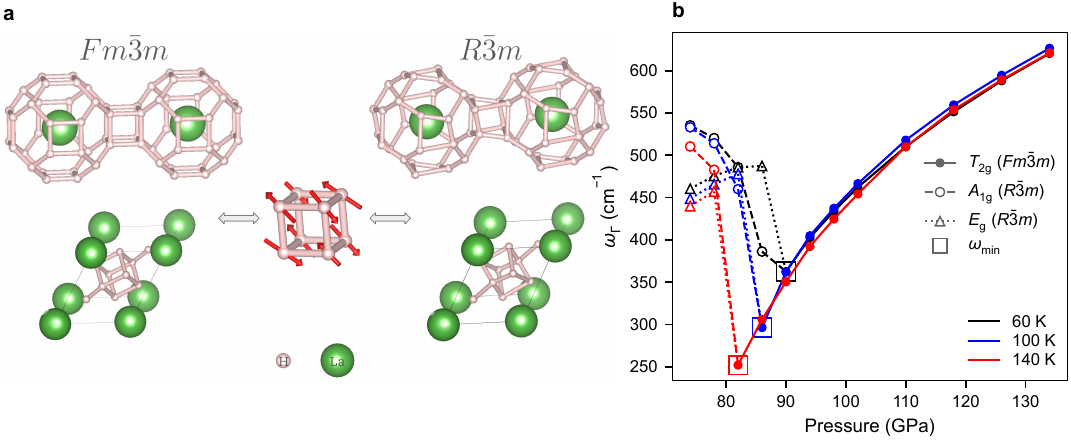}
\caption{\textbf{Structural transition and phonon softening.} \textbf{a,} $Fm\bar{3}m$ and $R\bar{3}m$ structures of LaH$_{10}$. The top row shows the overall cage structures, and the bottom row depicts the position of H cubes within the La sublattice. The force pattern corresponding to one of the 
three-fold degenerate 
soft optical phonons 
of the $Fm\bar{3}m$ lattice
driving this transition is shown in the middle. They exhibit $T_{2g}$ symmetry with characteristic shearing/rotational character.
$Fm\bar{3}m$ undergoes a rhombohedral distortion ($R\bar{3}m$) upon decompression. \textbf{b,} Frequency of the $T_{2g}$ soft optical phonons at $\Gamma$, three-fold degenerate, as a function of pressure at various temperatures, computed using displacement-displacement correlator from PIMD-NPT. At their softest frequency, they split into the $A_{1g}$ and $E_g$ modes of the $R\bar{3}m$ lattice, shown on the left-hand side of the frequency dip. Empty squares represent the transition pressure at various temperatures. 
}\label{lah10_structures}
\end{figure}

Our PIMD simulations show that the $Fm\bar{3}m$ structure undergoes a rhombohedral distortion ($R\bar{3}m$) upon decompression (Fig.~\ref{lah10_structures}a shows the $Fm\bar{3}m$ and $R\bar{3}m$ structures). This is consistent with a recent theoretical work \cite{2022KEV}, in which the authors performed PIMD simulations and used unit-cell parameters variation to 
detect
the phase boundary and the low-pressure distortion. However, it is worth noting that other theoretical/experimental reports vary, with some claiming that the low-pressure La sublattice is of $C2/m$ space group \cite{2021SUN, 2018HAN}, while others claim it is $R\bar{3}m$ \cite{2018GEB, 2020KRU}. $C2/m$ is a maximal subgroup and a slightly distorted version of $R\bar{3}m$. In the former, the three-fold rotation symmetry is broken, but inversion is preserved. The simulated powder XRD patterns of these two structures are quite similar~\cite{2022KEV}, and it is difficult to ascertain which one fits the experimental XRD pattern better.

Anharmonic phonons computed at $\Gamma$ 
are shown in Fig.~\ref{lah10_structures}b, which plots
the first optical phonon frequencies 
of the $Fm\bar{3}m$ lattice 
as a function of pressure at various temperatures 
(see Extended data Fig.~\ref{fm3m_nvt_phonon_dispersion} for full phonon dispersions). 
These optical phonon modes, three-fold degenerate, clearly soften as the pressure is lowered, then hit a floor at the transition, 
after which they jump discontinuously. 
Just after their minimum, they split into the $A_{1g}$ and $E_g$ modes of the $R\bar{3}m$ lattice, which
start hardening again. The minima in frequency enable us to trace a phase transition boundary. Moreover, the behavior of the soft phonons never going to zero provides the first hint that this transition is first-order. 
We identified the displacement pattern corresponding to the three-fold degenerate soft modes to have $T_{2g}$ symmetry. This could function effectively as an order parameter for this phase transition, and enables us to do a symmetry-based analysis within Landau's theory to provide a rigorous justification for the transition to be first-order. According to Landau's theory of phase transitions~\cite{1980LAN, 1987BIN}, the free energy expansion in terms of the order parameter should not contain a cubic term for the transition to be of second order. In other words, if the symmetric cubic product of the order parameter ($\Delta \otimes \Delta \otimes \Delta$) contains the irreducible representation $A_{1g}$, a cubic invariant exists, and the transition cannot be second order. $\Delta \otimes \Delta \otimes \Delta$ indeed contains $A_{1g}$\footnote{Using the standard product rules for the irreducible representation $T_{2g}$ in the $O_h$ point group:

\begin{equation}
    T_{2g} \otimes T_{2g} = A_{1g} \oplus E_{g} \oplus T_{1g} \oplus T_{2g}
\end{equation}
and then

\begin{align}
    T_{2g} \otimes T_{2g} \otimes T_{2g} &= (A_{1g} \oplus E_{g} \oplus T_{1g} \oplus T_{2g}) \otimes T_{2g} \\
    &= A_{1g} \oplus A_{2g} \oplus 2E_g \oplus 3T_{1g} \oplus 4T_{2g}
\end{align}}, and thus, a cubic term exists in the free energy expansion, clearly indicating it to be a first-order transition. 

Note, however, that the temperature dependence of the phonon frequency floor is not trivial. Indeed, the floor strongly decreases with temperature, suggesting that the energy barrier separating the two minima in the potential energy landscape (PES) is small, at least of the order of magnitude of room temperature fluctuations, giving the transition a weak first-order character (see Sec.~\ref{weak first-order} for more details).

\section{Quadrupolar order parameter}\label{quadrupolar_order}

\begin{figure}[H]
\centering
\includegraphics[width=1.0\textwidth]{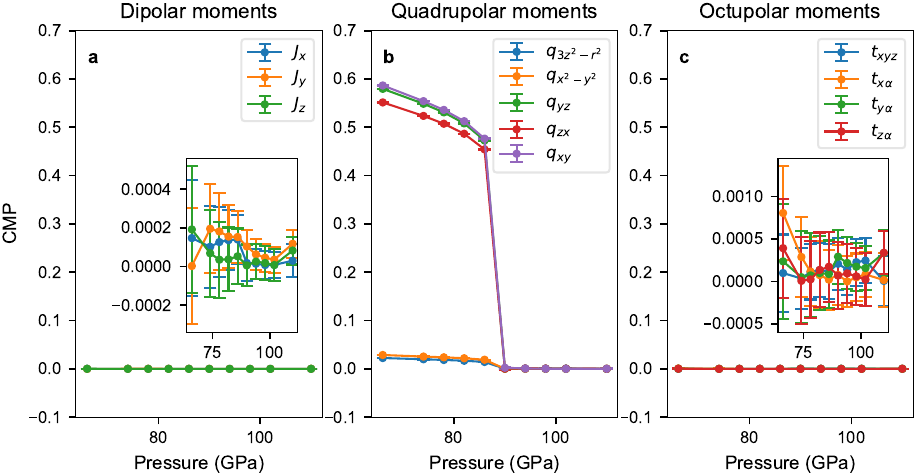}
\caption{\textbf{CMP moments averaged over H cubes and over the centroid trajectory.} \textbf{a.} Dipolar, \textbf{b.} quadrupolar and \textbf{c.} octupolar moments. Dipolar and octupolar moments are identically zero. Quadrupolar moments, however, are non-zero in the low-symmetry region, and zero in the high-symmetry region. Thus, they serve as a natural order parameter for this phase transition.}\label{cmp_moments}
\end{figure}

Although soft phonons indicate the onset of the structural instability, the order parameter yields the most rigorous and unambiguous characterization of the transition. It condensates the broken symmetry into a scalar quantity. To uncover the nature of symmetry breaking across the phase transition in LaH$_{10}$, we computed CMP moments for H cubes (see Fig.~\ref{lah10_structures}a) in the structure. At low pressures, the cubes distort, leading to various low-symmetry phases. Indeed, the geometry of the cubes (in terms of dihedral angles and bond lengths) can be used as a signature of different phases (see Extended data Fig.~\ref{dihedral_patterns} and Fig.~\ref{coutour_phase_diagram}). Fig.~\ref{cmp_moments} shows dipolar, quadrupolar, and octupolar moments computed for the cubes belonging to LaH$_{10}$ structures at various pressures (see Methods). Dipolar and octupolar moments are identically zero. Quadrupolar components, on the other hand, first decrease slowly at low pressures and then rapidly collapse to zero by rising the pressure, indicating a phase transition. The phases before and after the collapse of the quadrupolar moments are the low-symmetry (rhombohedral) $R\bar{3}m$ phase and the high-symmetry (cubic) $Fm\bar{3}m$ phase, respectively. The phases were identified by employing space group analysis \cite{2024TOG} of the time-averaged PIMD structures and also from the contour plots in the two-dimensional space of cube dihedral angles and corresponding H-H distance along the edges (see Extended data Fig.~\ref{coutour_phase_diagram}). Dipolar and octupolar moments being identically zero can be understood from the fact that the distortion from $Fm\bar{3}m$ to $R\bar{3}m$ is centrosymmetric in nature, namely the distribution of atoms in the cubes is centrosymmetric with respect to its center. The point group of the cubes in the high-symmetry phase and distorted cubes in the low-symmetry phase are $O_h$ and $D_{3d}$, respectively, which indeed both have an inversion center.

\begin{figure}[H]
\centering
\includegraphics[width=1.0\textwidth]{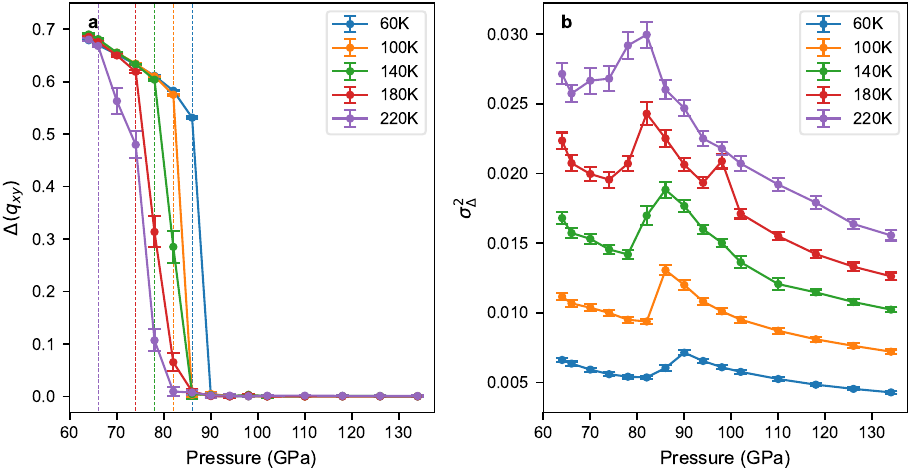}
\caption{\textbf{Pressure evolution of the global order parameter across the phase transition.} \textbf{a,} Order parameter as a function of pressure at several temperatures. The transition to the high-symmetry phase is defined to occur at the pressure at which the order parameter starts to collapse (dotted line). \textbf{b}, Variance of the order parameter under the same conditions. Fluctuations peak near the transition.}\label{order_param_and_var}
\end{figure}

Quadrupolar moments thus serve as a natural order parameter for this phase transition. 
In LaH$_{10}$ this corresponds to the basis of \textit{d} ($l=2$) spherical harmonics (see Sec.~\ref{sec11.1}). In the $O_h$ point group, they admit $T_{2g}$ and $E_g$ symmetries as irreducible representations. Moreover, the $T_{2g}$ components ($q_{xy}$, $q_{yz}$ and $q_{zx}$) give the maximum signal in the low symmetry phase, the $E_g$ components being close to zero (see Fig.~\ref{cmp_moments}). The three $T_{2g}$ components satisfy $q_{xy}\simeq q_{yz}\simeq q_{zx}$, indicating condensation along the (1,1,1) direction in the $T_{2g}$ order parameter space. This is consistent with the symmetry reduction $O_h\rightarrow D_{3d}$. This makes $T_{2g}$ the primary order parameter of this phase transition, and is consistent with the phonon picture where the soft-mode phonons indeed have a $T_{2g}$ symmetry (see Fig.~\ref{lah10_structures}b).
Based on this symmetry analysis, we can precisely locate the symmetry-breaking transition from our PIMD simulations by scanning the samples drawn from the quantum partition function via the quantities depicted in Fig.~\ref{cmp_moments}. 
Phase transition towards the high-symmetry phase is defined to occur at the pressure where the $T_{2g}$ order parameter starts to collapse, which for higher temperatures occurs at lower and lower pressures (Fig.~\ref{order_param_and_var}(a)). Phase transition also corresponds with a peak in the variance of the order parameter (Fig.~\ref{order_param_and_var}(b)), whose fluctuations are the highest near the transition point.  

\begin{figure}[H]
\centering
\includegraphics[width=0.8\textwidth]{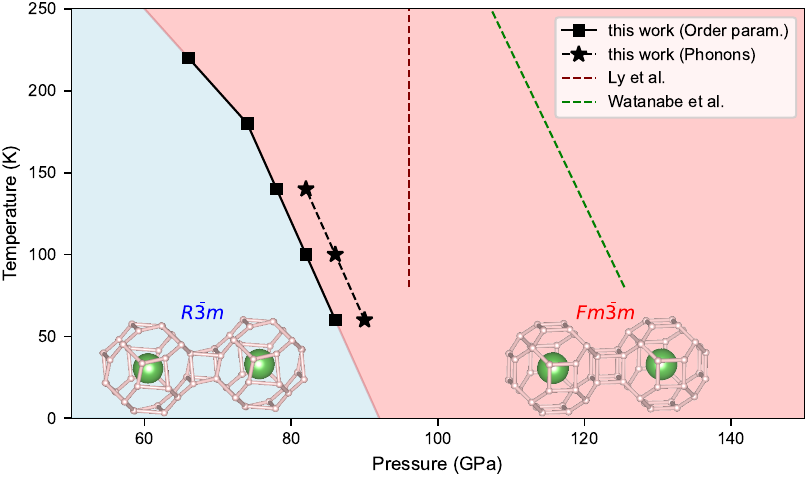}
\caption{\textbf{$P-T$ phase diagram of LaH$_{10}$.} The phase boundary identified using the soft phonon agrees with the one identified using the collapse of the order parameter, showing that the two methods are consistent. Phase transitions as determined in previous reports are also shown for comparison~\cite{2022KEV, 2022WAT}. Ly et al.~\cite{2022KEV} identified the low-pressure distortion as $R\bar{3}m$ and placed the transition at $\approx $96 GPa, which is close to the current work. It should be noted that in both these works, calculations were done over a much coarser $P$-$T$ grid, making it difficult to determine a precise phase boundary.}\label{quantum_phase_diagram}
\end{figure}

\section{Location of the quantum transition line}\label{phase_diagram}
The phase boundary of the symmetry-breaking transition is shown in Fig.~\ref{quantum_phase_diagram}, where we trace two independent signatures: the collapse of the order parameter and the jump in the soft phonon frequency. The phonon and order parameter pictures are mutually consistent. Nuclear quantum fluctuations shift the phase transition line to lower pressures as compared to classical simulations, and stabilize $Fm\bar{3}m$ up to pressures as low as $\sim$ 80-90 GPa. We further compare the phase transition boundary determined in this work with previous reports. Our transition line is quite close to the one determined by Ly et al.~\cite{2022KEV}, who also used PIMD simulations to find the equilibrium structure at various P-T conditions. However, they did calculations over a coarser $P$-$T$ grid and could not resolve the temperature dependence of the phase boundary. Moreover, the one reported by Watanabe et al.~\cite{2022WAT} deviates from that obtained in the present study by approximately 40 GPa. This discrepancy can likely be ascribed to the shorter time length of their PIMD
simulations, the smaller number of beads used resulting in an incomplete treatment of NQEs, and the use of a relatively coarse pressure–temperature grid in their calculations. Experimentally, the phase transition is reported to occur at 138 GPa \cite{2021SUN}, and hence the theoretical quantum phase transition line overestimates the $Fm\bar{3}m$ stability region by about $\sim$ 50--60 GPa. A well-converged PIMD simulation provides an almost exact inclusion of nuclear quantum fluctuations; thus, this difference should be due to the incomplete treatment of electronic exchange-correlation effects by the PBE functional,
used in all these works.

\section{Weak first-order character}\label{weak first-order}
\begin{figure}[H]
\centering
\includegraphics[width=1.0\textwidth]{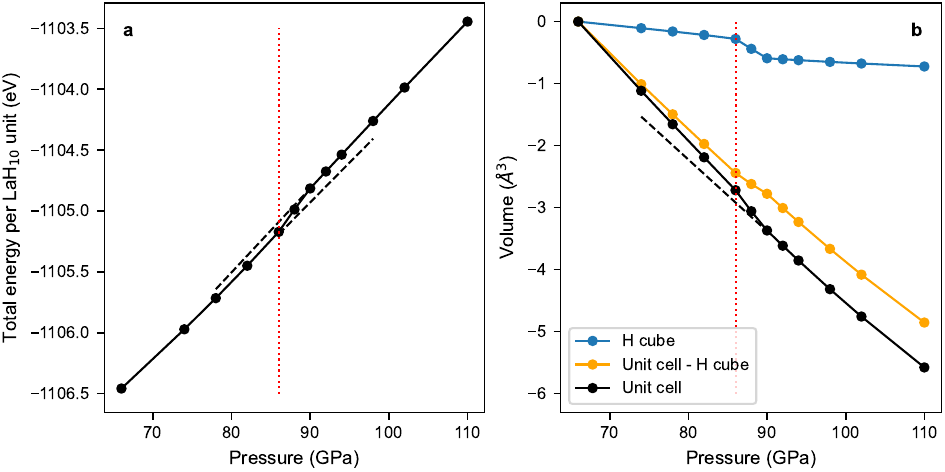}
\caption{\textbf{Equation of state across phase transition obtained from PIMD-NPT simulations at 45 K using a $4\times4\times4$ supercell.} \textbf{a,} Total energy per LaH$_{10}$ unit as a function of pressure. Black dashed lines depict the extrapolation of the high-symmetry phase into the low-pressure regime and of the low-symmetry phase into the high-pressure regime. \textbf{b,} Unit-cell volume (relative to the value at 66 GPa) as a function of pressure.  Both $E(P)$ and $V(P)$ exhibit two successive slope changes (red dotted line marks the onset), bounding a finite transition region consistent with a weak first-order transition. }\label{eos}
\end{figure}

Group theory analysis within Landau's framework provides rigorous evidence for the transition to be first-order; however, from our phonon analysis, the first-order character of the transition gets weaker and weaker 
as temperature rises. We confirm these features from thermodynamic evidence. There is a small discontinuity in the total energy 
per LaH$_{10}$ unit
at the phase transition (Fig.~\ref{eos}(a)), consistent with a weakly first-order character already at a temperature of 45 K. The discontinuity is also visible in the unit-cell volume near the transition pressure (Fig.~\ref{eos}(b)). Phase coexistence region, although narrow, is also visible in both plots. The jump is indeed very tiny, with the $Fm\bar{3}m$ phase being $\approx$ 0.1 eV higher than the $R\bar{3}m$ phase in energy at the transition pressure. In terms of cell volume, there is a slight contraction ($\approx$ 0.2 \AA{$^3$}) from the low-symmetry to the high-symmetry phase at the same pressure. Interestingly, if we consider the volume of H cubes only, the contraction is much more dramatic, but it is associated with an expansion in the remaining volume. $\frac{\Delta V}{V}$ at the transition pressure is very low, about 0.5 \%.
The smallness of $\frac{\Delta V}{V}$ implies a small latent heat, thus making the transition highly susceptible to thermal fluctuations, which could appear continuous in experiments and elusive in thermodynamic detection.

\begin{figure}[H]
\centering
\includegraphics[width=0.6\textwidth]{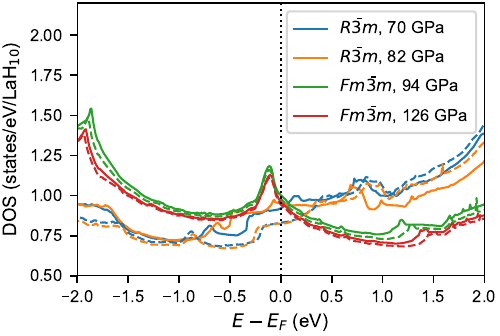}
\caption{\textbf{Electronic DOS at different pressures across the transition.} Solid lines show electronic DOS for quantum centroid structures, while dotted lines indicate DOS for classically relaxed structures. A Van Hove singularity is visible just below the Fermi energy in the DOS of the $Fm\bar{3}m$ phase.}
\label{electronic_dos}
\end{figure}
 This weak first-order nature is evident again when we compute other descriptors such as the electronic density of states (DOS) at the Fermi level (Fig.~\ref{electronic_dos}) across the transition. There is a Van Hove singularity just below the Fermi level in the high-symmetry $Fm\bar{3}m$ phase, which is consistent with previous reports \cite{2020ERR}. At the Fermi level, however, there is almost no change in the DOS for the two phases across the phase transition when quantum centroids are considered for structures. Moreover, there is a negligible change in DOS between the quantum centroid structure and the classically relaxed structure for $Fm\bar{3}m$. However, for $R\bar{3}m$, the two DOS are noticeably different, possibly because this low-symmetry phase is strongly renormalized by quantum effects. In $Fm\bar{3}m$, on the other hand, vibrations are more constrained due to high symmetry. The marginally small variations in physical quantities such as DOS is another manifestation that the first-order character of this transition is inherently weak. The DOS value is naturally related to the critical temperature estimate in the superconducting phase. Despite this small DOS variation across the structural phase transition, the measured $T_c$ shows a quite steep increase~\cite{2021SUN} in the same pressure range, which should be attributed to changes in the phonon behavior.

\section{Quadrupolar phonons and superconductivity}\label{phonons_and_superconductivity}
It is generally believed that conventional hydride superconductors would have high electronic DOS at the Fermi energy ($E_F$), possibly enhanced by a Van Hove singularity. However, there are counterexamples to this. For instance, $Im\bar{3}m$ CaH$_6$ and $Fm\bar{3}m$ ThH$_{10}$ actually have a dip structure in DOS around $E_F$ \cite{2024DAR}. In LaH$_{10}$, although we do see a peak in DOS slightly below $E_F$ (Fig.~\ref{electronic_dos}), the change in DOS at $E_F$ before and after the phase transition is negligible and cannot explain the steep variation of the measured $T_c$ across the structural transition. Hence, the coupling of quadrupolar soft-mode phonons, which play a pivotal role in driving the symmetry-breaking transition, should also be integral for superconductivity in this material.

To shed light on their role,
it is interesting to compare the phonon softening in LaH$_{10}$ with H$_3$S, another hydride superconductor. H$_3$S is characterized by a second-order displacive transition between the centrosymmetric $Im\bar{3}m$ and the polar $R3m$ phase~\cite{2017GON, 2020MIN, 2026CHE}, whose order parameter is dipolar and associated with the H position with respect to the two neighboring S atoms \cite{2026CHE}. The dipolar nature of this transition is clearly different from the quadrupolar one in LaH$_{10}$. Moreover, it has been shown that in H$_3$S, 
optical phonons at $\Gamma$ vanish as 
the transition is approached
from the high-symmetry side,
with a jump of their frequencies in the polar phase. This behavior is explained by the vanishing curvature of the PES at the phase boundary, typical of second-order transitions. 
In LaH$_{10}$ the presence of soft $T_{2g}$ optical phonons at $\Gamma$, that harden upon decreasing temperature instead of vanishing, implies that the PES always has a finite curvature as opposed to H$_3$S, yielding the different order of the transition,
as revealed in the previous Sections.
The weak first-order character of the quadrupolar transition in LaH$_{10}$ could have non-trivial consequences for its superconductivity.
In H$_{3}$S, the first three optical modes, that soften at the displacive transition,~\cite{2026CHE} do not carry a significant electron-phonon coupling, as rationalized by Fr\"ohlich's mechanism, 
which excludes transverse phonons from the coupling term~\cite{1950FRO}. The major contribution to the electron-phonon coupling comes from high-frequency phonon modes \cite{2015ERR}. 
However, in LaH$_{10}$ the soft phonons at $\Gamma$ are quadrupolar,  
they go beyond Fr\"ohlich's theory, and
their coupling with the electronic states 
is allowed in the long-wavelength limit~\cite{brunin2020electron,jhalani2020piezoelectric}. 
In fact, there is a peak in $\alpha^2F(\omega)$ for these
low-frequency phonons as computed in Refs.~\cite{2020ERR, 2020KRU, 2019WAN}, signaling that their contribution to the total electron-phonon coupling $\lambda$ cannot be neglected. Moreover, the weakness of the first-order character, that makes these modes soft, can enhance the superconducting $T_c$, because their contribution to $\lambda$ scales as $1/\omega^2_\textrm{soft}$, with
$\omega_\textrm{soft}$ 
being
their small frequency value in the proximity of the transition. 
The electron-phonon coupling enhancement by soft phonons is general and can be found in other superhydrides~\cite{2022FRA}.

\section{Conclusions}\label{sec13}

We have characterized the phase transition in LaH$_{10}$ from the rhombohedral ($R\bar{3}m$) to the high-symmetry cubic ($Fm\bar{3}m$) structure. We determined a precise boundary between the two phases by a full quantum treatment of nuclei, using PIMD simulations. We employed a cluster multipole moments analysis, applied to the H-cubes intercalating the La sublattice, in order to characterize the order parameter of the transition.  We discovered that its nature is quadrupolar with $T_{2g}$ symmetry. Based on Landau's theory, phonon softening, and thermodynamic properties, we identified this phase transition to be weak first-order.
Finally, we highlighted the importance of quadrupolar phonon softening for sustaining high-temperature superconductivity in this material. Multipolar order parameters can be found in the structural transitions of other superhydride superconductors, owing to their richness in H content. Following the LaH$_{10}$ paradigmatic example, quadrupolar distortions of the hydrogen sublattice can be a key ingredient, not fully explored yet, in stabilizing high-temperature superconductivity near structural phase transitions.
Thus, our findings identify proximity to weak first-order multipolar transitions as a guide in the search for new high-$T_c$ superconductors in hydrogen-rich compounds.

\section{Methods}\label{sec11}
\subsection{Generating machine learning interatomic potential}\label{mlp}
To obtain equilibrium crystal structures including NQEs for the H atoms, we performed 
\emph{ab initio}
PIMD simulations~\cite{1993TUC, 1996DOM, 2006MOT}. Phase space exploration can require MD trajectories of 50–100 ps or even longer, depending on the complexity of the free energy landscape. Furthermore, at low temperatures, quantum fluctuations can be quite dominant, and PIMD calculations can require $\approx$ 100 beads for their proper treatment, making these calculations prohibitively expensive. Hence, we generate and utilize a MACE (message-passing atomic cluster expansion) MLIP to speed up PIMD calculations. MACE is an equivariant message-passing neural network interatomic potential~\cite{2022BAT, 2023KAV} that constructs higher body-order features following the atomic cluster expansion. Our model uses 2 message-passing layers, with a cutoff radius of 5 \AA{}. Each atomic feature vector comprises of 128 rotationally invariant (scalar) and 128 equivariant vector channels (hidden irreps 128$\times$0e + 128$\times$1o).

For training the MACE model, configurations sampled from classical MD (NVT ensemble) were utilized. These MD simulations were performed at 250K and for a number of cell volumes. \emph{Ab initio} MD for training was performed on $2\times2\times2$ supercells (88 atoms) of LaH$_{10}$, and trajectories were generated considering 4 different starting structures (the space groups $Fm\bar{3}m$, $R\bar{3}m$, $Immm$, and $C2$). Classically relaxed structures at pressures ranging from 60 GPa to 220 GPa were used as starting configurations for the MD simulations. A Langevin thermostat was employed to maintain the simulation temperature, with a time step of 0.8 fs used for integration. The total simulated time was $\approx$12 - 16 ps for each of these simulations. The energies and forces for MD were calculated at the DFT level using the GGA-PBE exchange-correlation functional~\cite{1996PER} and the quantum ESPRESSO code~\cite{2009GIA, 2017GIA}. The pseudopotentials used were norm-conserving and scalar-relativistic, produced using the ONCVPSP code (Optimized Norm-Conserving Vanderbilt Pseudopotential)~\cite{VANSETTEN201839, 2013HAM} with 11 valence electrons for La and a single valence electron for H. The cutoff energies for plane-wave expansion and charge density were chosen to be 100 Ry and 800 Ry, respectively, and a $\Gamma$-centered $9\times9\times9$ Monkhorst-Pack~\cite{1976MON} k-point grid for Brillouin zone integration. Finally, a total of 9300 configurations, selected at random from the MD trajectories, were used for generating the MLIP, using a learning rate of 0.01. Of these, 7440 configurations were used in the training set, and the remaining 1860 configurations formed the test set. The trained model had an RMSE (computed on the test set) of 0.2 meV/atom for energies and 20.7 meV/\AA{} for forces (see Extended data Fig.~\ref{model_training}, Fig.~\ref{model_eval} and 
Fig.~\ref{model_benchmark} for training convergence, evaluation of the model on the test set, and model validation, respectively).

\subsection{PIMD and anharmonic phonons}\label{pimd}
Using the trained MLIP, production PIMD calculations were performed on $3 \times 3 \times 3$ supercells, using the i-Pi code~\cite{2019KAP, 2024KAP}. These calculations were run using the NPT ensemble for a simulation length of 64 ps. A time step of 0.4 fs was used for MD integration. For temperature control, we used the path integral Langevin equation (PILE) thermostat~\cite{2010CER}, which attaches a local Langevin thermostat to the centroid mode. For pressure control, a flexible barostat was used, allowing for full cell fluctuations. For the lowest temperature simulation (60 K), 76 beads were used, and for other temperatures, the number of beads 
($N_\textrm{beads}$)
was scaled inversely proportional to temperature, keeping the product 
$N_\textrm{beads} \times T$
constant, ensuring consistent convergence of the path integral.

Phonon spectra provide a direct fingerprint of structural instabilities, with soft modes signaling the onset of symmetry-breaking transitions. We computed anharmonic phonons incorporating NQEs from the zero-time Kubo-transformed displacement–displacement correlators, directly accessible from the PIMD centroid trajectories following the approach of Morresi et al.~\cite{2021MOR}

\subsection{Devising the order parameter}\label{sec11.1}

To define the order parameter, we start by computing cluster multipole (CMP) moments~\cite{2017SUZ} for H cubes. The CMP rank-p moment for the $\mu$th H cube is defined as follows:
\begin{equation}
    M_{pq}^{(\mu)} \equiv \sqrt{\frac{4\pi}{2p+1}} \sum_{i=1}^{N_{\text{atom}}^{(\mu)}} \boldsymbol{R}_i \cdot \nabla_i (|\boldsymbol{R}_i|^p Y_{pq} (\theta_i, \phi_i)^*),
\end{equation}
where $N_{atom}^{\mu} = 8$ is the number of atoms in the $\mu$th cube, $\boldsymbol{R}_i \equiv (X_i, Y_i, Z_i)$ is the displacement vector of the $i$th atom in the cube from the center of the cube-fragment, $\nabla_i \equiv \frac{\partial}{\partial \boldsymbol{R}_i}$, $Y_{pq}$ are the spherical harmonics 
of degree $p$ and order $q$,
and $\theta_i$, $\phi_i$ are the polar angle and azimuthal angle respectively. The center of a cube is defined as the center of the best-fitting sphere, fitted to the points of the cube using least-squares fitting. Due to the cubic symmetry of the high symmetry $Fm\bar{3}m$ structure, the CMP moments are classified according to the irreducible representations of the $O_h$ point group. The classification is provided in Ref. \cite{2017SUZ}. CMP moments up to rank 3 were computed, giving dipole, quadrupole, and octupole moments. For instance, the quadrupolar CMP moments are defined as follows:
\begin{align}
    q_{3z^2-r^2} &\equiv M_{20} \\
    q_{x^2-y^2} &\equiv \frac{1}{\sqrt{2}} (M_{22} + M_{2-2}) \\
    q_{yz} &\equiv -\frac{i}{\sqrt{2}} (M_{21} + M_{2-1}) \\
    q_{zx} &\equiv \frac{1}{\sqrt{2}} (-M_{21} + M_{2-1}) \\
    q_{xy} &\equiv \frac{i}{\sqrt{2}} (M_{22} - M_{2-2})
\end{align}
To quantify the global structural order, the CMP moments are spatially averaged over all $N_{\text{cube}}$ cubes at each time step $t$, and the resulting quantity is averaged over the MD trajectory.
\begin{equation}
    \Delta_l = \left\langle \frac{1}{N_{\text{cube}}} \sum_{\mu=1}^{N_{\text{cube}}} O_l^{(\mu)}(t) \right\rangle_{t},
\end{equation}
where, $\Delta_l$ is the global order parameter, $O_l^{(\mu)}(t)$ denotes the relevant CMP moment of symmetry species $l$ for the $\mu$th cube at time $t$. To characterize the thermal fluctuations of the order parameter, we further compute its variance, which serves as a susceptibility-like diagnostic. A peak in susceptibility as a function of pressure indicates enhanced fluctuations near a structural phase transition, analogous to the divergence of the order parameter susceptibility at a critical point.

\backmatter

\bmhead{Supplementary information}

The data supporting the findings of this article are provided in the Extended Data section.

\bmhead{Acknowledgements}
The authors acknowledge GENCI for providing computational resources on the IDRIS Jean-Zay supercomputing clusters and TGCC Joliot-Curie Rome partition under project numbers A0170906493 and A0190906493. We would also like to thank European High Performance Computing (EuroHPC) for the computational grant EHPC-EXT-2024E01-064 allocated on Leonardo (booster partition). We are grateful to EPICURE, a EuroHPC Joint Undertaking (JU) initiative, for supporting this project on Leonardo through the EuroHPC JU 2024E01 call for proposals for extreme scale access mode. We also thank the EuroHPC-JU for the support through the ``EU-Japan Alliance in HPC'' HANAMI project (HPC Alliance for Applications and supercomputing Innovation: the Europe - Japan collaboration. This work was also supported by Grants-in-Aid for Scientific Research from JSPS (KAKENHI Grants No. 25H01252 and No. 24H00190), JST K-Program JPMJKP25Z7, and the RIKEN TRIP initiative (RIKEN Quantum, Advanced General Intelligence for Science Program, Many-body Electron Systems). K.N. acknowledges financial support from MEXT Leading Initiative for Excellent Young Researchers (Grant No.~JPMXS0320220025). A.R. and K.N. acknowledge financial support from JST BOOST (Grant No.~JPMJBY24F3).

\section*{Declarations}

\begin{itemize}
\item Conflict of interest/Competing interests: The authors declare no conflict of interest/competing interests.
\item Data availability: Supporting information is provided in the extended data. All other data is available from the corresponding authors upon reasonable request.
\item Code availability: Quantum ESPRESSO used for DFT calculations is open-source and available at \href{https://www.quantum-espresso.org/}{https://www.quantum-espresso.org/}. MACE and i-Pi, used to generate MLIP and perform PIMD, respectively, are also both open-source and available at \href{https://github.com/acesuit/mace}{https://github.com/acesuit/mace} and \href{https://ipi-code.org/}{https://ipi-code.org/}. The code used to compute anharmonic phonons from PIMD is an in-house code developed by some of the authors. The method used is well-documented in the scientific literature.
\end{itemize}

\noindent

\begin{appendices}

\section{Extended data}\label{secA1}

\renewcommand{\theHfigure}{extfig.\arabic{figure}}
\renewcommand{\theHtable}{exttab.\arabic{table}}
\renewcommand{\figurename}{Extended Data Fig.}
\renewcommand{\thefigure}{\arabic{figure}}
\renewcommand{\theHfigure}{extfig.\arabic{figure}}
\renewcommand{\tablename}{Extended Data Table}
\renewcommand{\thetable}{\arabic{table}}
\renewcommand{\theHtable}{exttab.\arabic{table}}

\begin{figure}[H]
\centering
\includegraphics[width=0.6\textwidth]{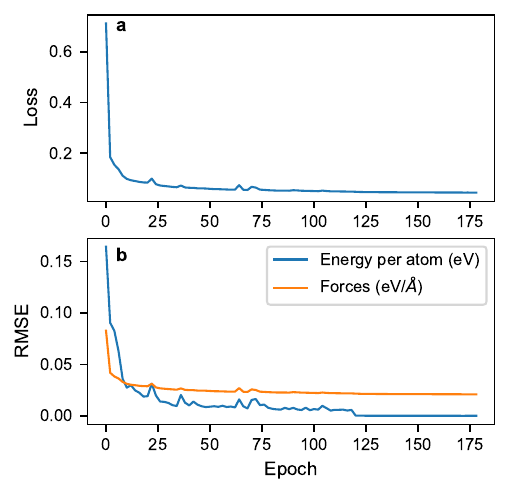}
\caption{\textbf{Two-stage training convergence of the MACE potential.} \textbf{a,} Combined loss versus training epoch and \textbf{b,} Validation RMSE of energies and forces versus training epoch. Energy and force terms in the combined loss are weighted 1:100 for the first 120 epochs and 10:1 for the remaining 60, accounting for the discontinuity at epoch 120. Since there are $3N$ force components per configuration and only 1 energy value, the energy signal is weaker. Thus, it helps to have a much higher weight for forces for most of the training process to learn the geometry of the PES and then later fine-tune it to get correct energies.
}\label{model_training}
\end{figure}

\begin{figure}[H]
\centering
\includegraphics[width=1.0\textwidth]{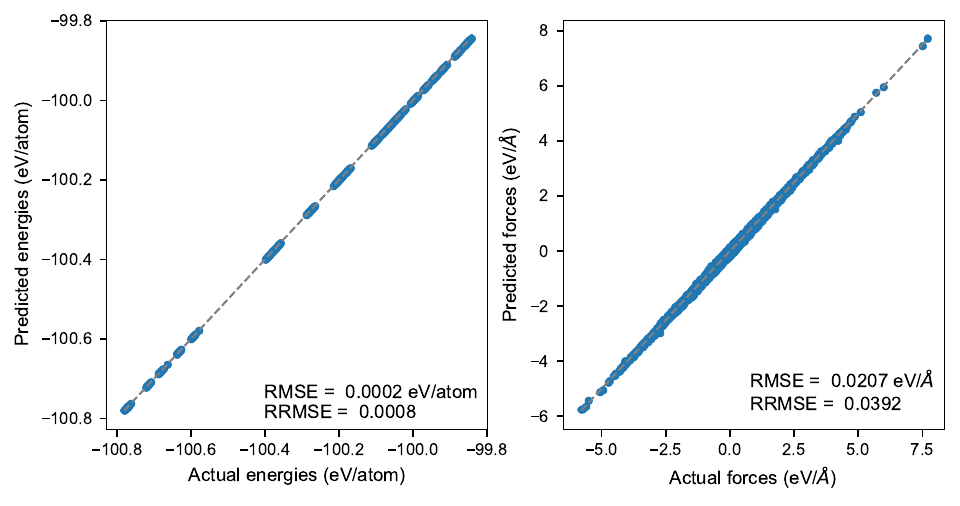} 
\caption{\textbf{Performance of the MLIP model on the test set}. Scatter plots compare predicted versus actual energies (left) and forces (right). The RMSE quantifies the model error on the test set, i.e., how closely the predicted values match the reference ones. The relative RMSE (RRMSE) is the RMSE normalized by the standard deviation of the reference values in the test set, so that $\mathrm{RRMSE} < 1$ indicates predictive power beyond that of a mean predictor. RMSEs in energies and forces over the test set were 0.2 meV/atom and 20.7 meV/\AA{} respectively.
}\label{model_eval}
\end{figure}

\begin{figure}[H]
\centering
\includegraphics[width=1.0\textwidth]{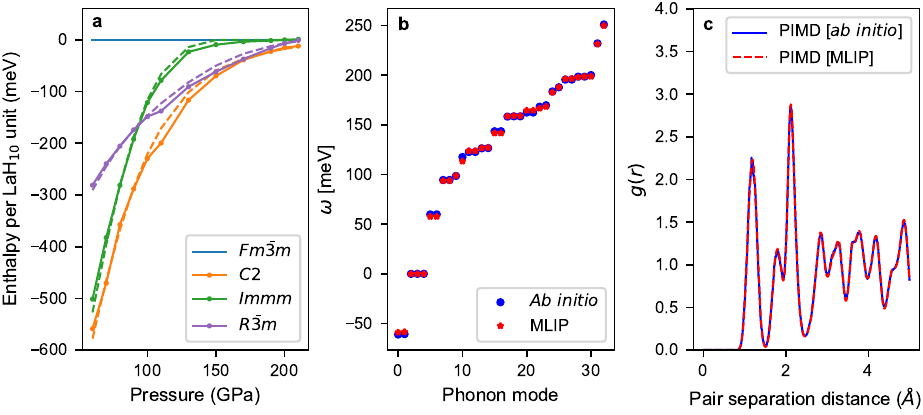}
\caption{\textbf{Validation of the MLIP against three important benchmarks each probing a different feature of the PES.} \textbf{a,} MLIP predicted enthalpies (dashed lines) at 0 K compared with \emph{ab initio} values (solid lines) to assess the accuracy of total energies. \textbf{b,} Frozen phonons (supercell method) at $\Gamma$ computed for the $R\bar{3}m$ phase at 130 GPa to probe the local curvature of the PES around equilibrium. \textbf{c,} Radial distribution function ($g(r)$) computed for PIMD-NVT simulations at 248 K to assess the finite-temperature behaviour in the 
quantum nuclear
regime. These
tests validate the performance of MLIP not only on static energetics but also on the
dynamical and structural properties.
}\label{model_benchmark}
\end{figure}

\begin{figure}[H]
\centering
\includegraphics[width=0.9\textwidth]{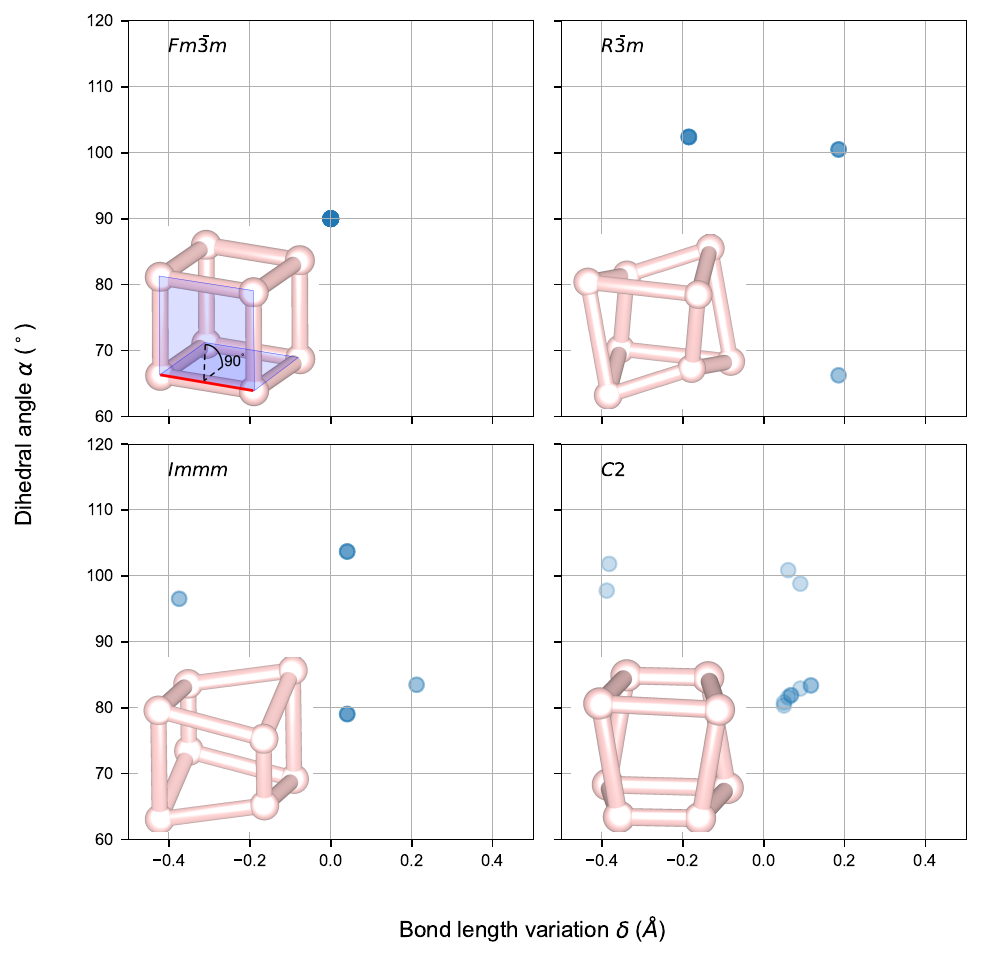}
\caption{\textbf{Dihedral angle 
$\alpha$
versus H–H bond length 
variation $\delta$
plots for the hydrogen cubes serve as distinct signatures for distinguishing the different static structures of LaH$_{10}$.} Each dihedral angle is the angle between two adjacent planes of a cube, and the corresponding bond is their common edge; the 
bond length 
variation $\delta$
is obtained by subtracting the mean bond length of each cube from its individual bond lengths. Darker points indicate a higher degeneracy due to overlapping data. The associated H-cube structures are shown as insets, with the $Fm\bar{3}m$ inset illustrating the dihedral angle between two planes and its corresponding bond. The distinct patterns serve as signatures for distinguishing the different phases.
}
\label{dihedral_patterns}
\end{figure}

\begin{figure}[H]
\centering
\includegraphics[width=1.\textwidth]{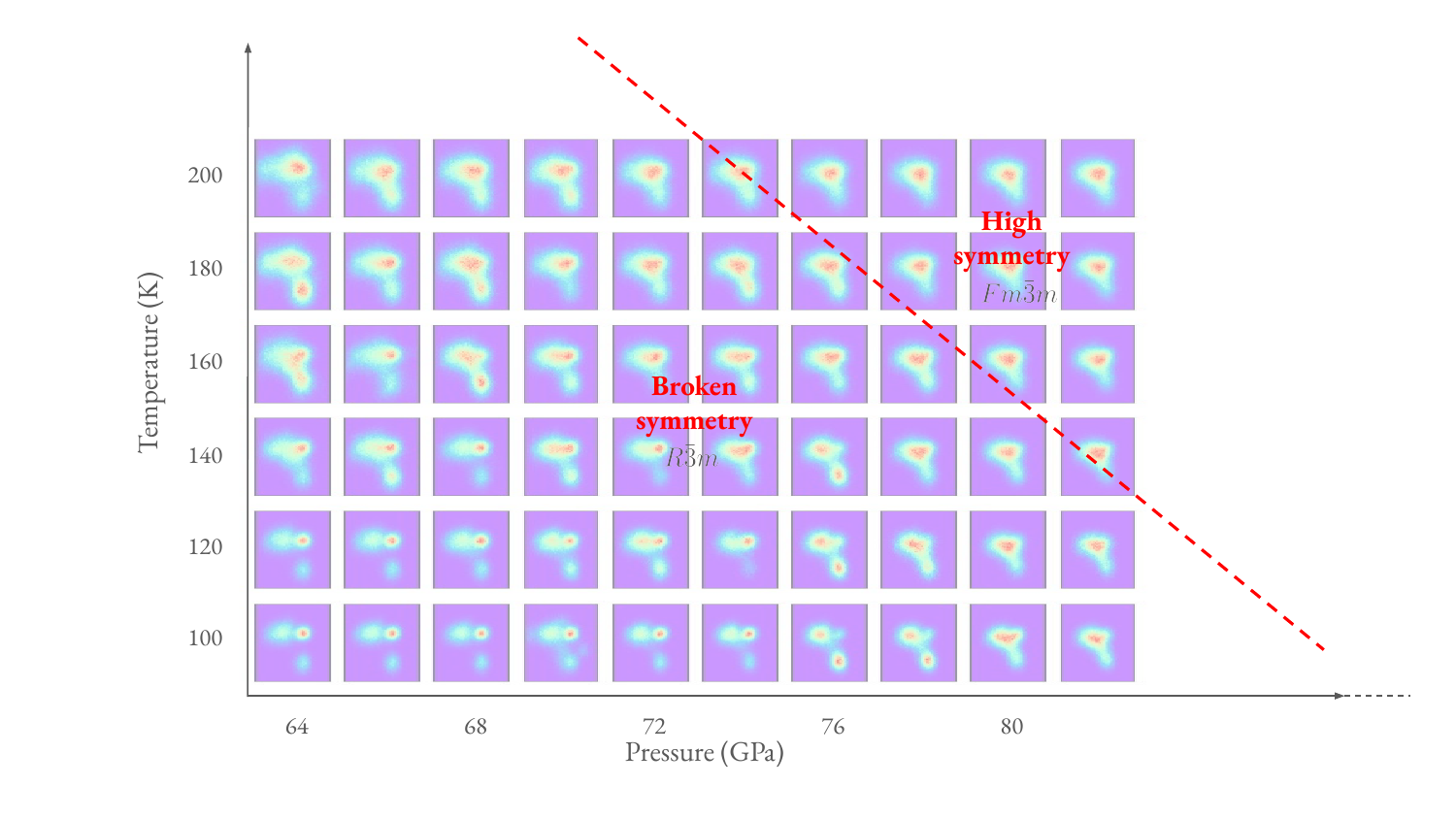}
\caption{\textbf{Symmetrization of the hydrogen sublattice with increasing pressure and temperature mapped by contour plots of H-cube dihedral angles and the corresponding H-H bond length 
variations
.} From PIMD trajectories, we can construct heat maps of dihedral angles versus bond length
variations
for different P-T conditions. The scatter plots for static structures (Fig.~\ref{dihedral_patterns}) turn into density distributions. These distributions have distinct shapes, which correspond to the scatter plots of the static structures. For instance, $Fm\bar{3}m$ being a single degenerate static point at (0, 90), becomes a distribution with one compact bright peak, while the low-symmetry $R\bar{3}m$ spreads into a characteristic three-lobed pattern. The red dotted line marks an approximate phase-transition boundary.}
\label{coutour_phase_diagram}
\end{figure}

\begin{figure}[H]
\centering
\includegraphics[width=0.6\textwidth]{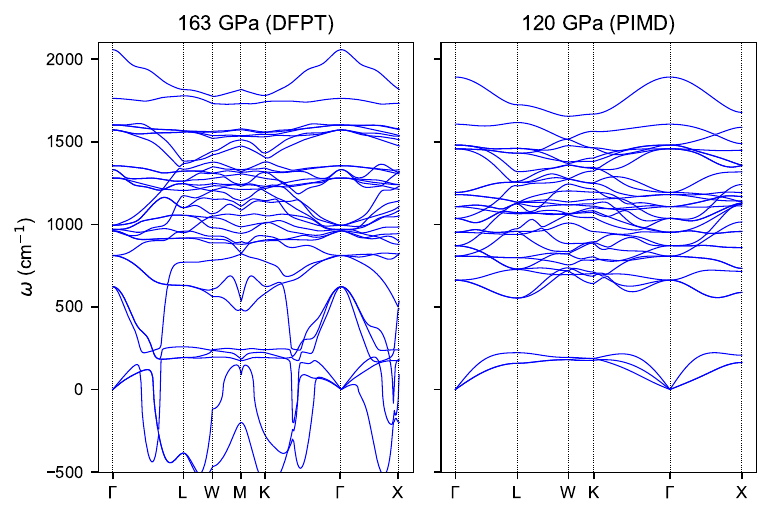}
\caption{\textbf{Harmonic phonons computed using density functional perturbation theory compared with anharmonic phonons computed from PIMD-NVT trajectory (T = 100 K, 32 beads, 32 ps simulation), for the $Fm\bar{3}m$ structure.} Harmonic phonons depict $Fm\bar{3}m$ to be dynamically unstable at 163 GPa; anharmonic phonons, on the other hand, predict this structure to be dynamically stable at much lower pressures.}

\label{dfpt_vs_pimd_phonons}
\end{figure}

\begin{figure}[H]
\centering
\includegraphics[width=1.0\textwidth]{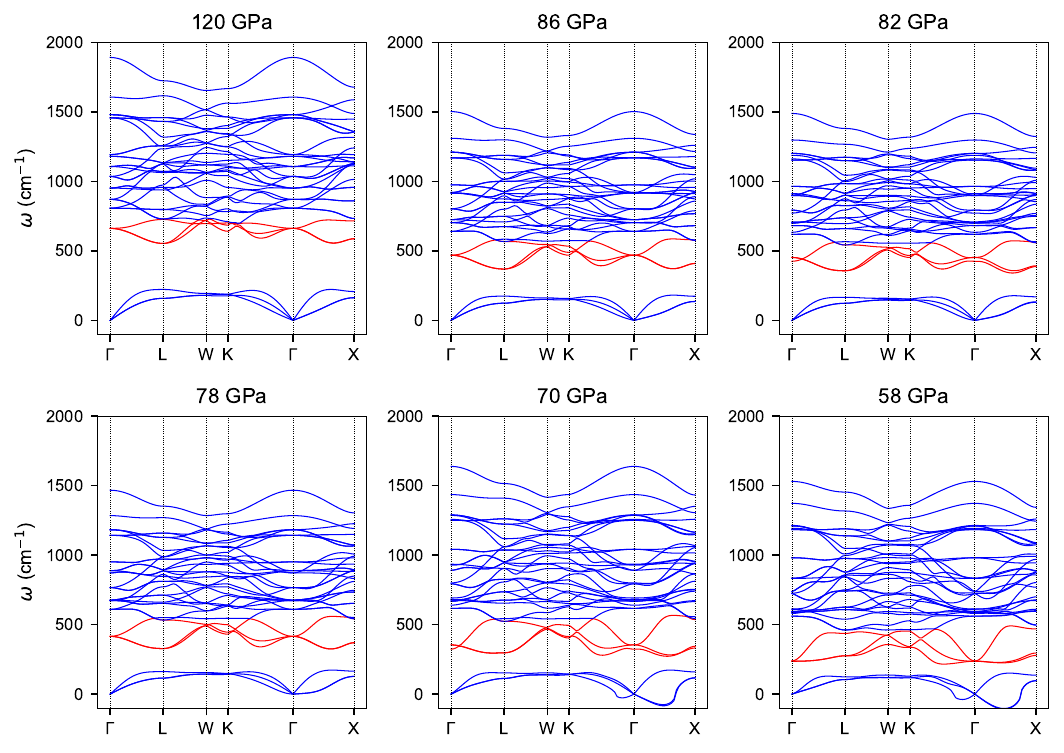}
\caption{\textbf{Anharmonic phonon band structure of $Fm\bar{3}m$ LaH$_{10}$ at different pressures, computed from PIMD-NVT trajectories (T = 100 K, 32 beads, 32 ps simulation).} We see phonon softening as the pressure is lowered. The $Fm\bar{3}m$ structure 
cannot
physically transform into the low-symmetry $R\bar{3}m$ structure in these simulations, because the lattice vectors are frozen due to the NVT nature of these simulations. Still, it shows a clear and systematic softening of the lowest optical branches (highlighted in red), which are precisely the modes that drive the $Fm\bar{3}m \rightarrow R\bar{3}m$ distortion.}

\label{fm3m_nvt_phonon_dispersion}
\end{figure}

\end{appendices}

\bibliography{sn-bibliography}

\end{document}